\documentclass[sigconf]{acmart}

\AtBeginDocument{%
  \providecommand\BibTeX{{%
    \normalfont B\kern-0.5em{\scshape i\kern-0.25em b}\kern-0.8em\TeX}}}

\setcopyright{acmcopyright}

\copyrightyear{2022}
\acmYear{2022}
\acmConference[CEP 2022]{Computing Education Practice 2022}{January 6, 2022}{Durham, United Kingdom}
\acmBooktitle{Computing Education Practice 2022 (CEP 2022), January 6, 2022, Durham, United Kingdom}

\begin{document}

\title{Feedback and Engagement on an Introductory Programming Module}

\author{Beate Grawemeyer}
\email{Beate.Grawemeyer@coventry.ac.uk}
\affiliation{%
  \institution{Coventry University}
  \city{Coventry}
  \country{U.K.}
}

\author{John Halloran}
\email{John.Halloran@coventry.ac.uk}
\affiliation{%
  \institution{Coventry University}
  \city{Coventry}
  \country{U.K.}
}

\author{Matthew England}
\email{Matthew.England@coventry.ac.uk}
\affiliation{%
  \institution{Coventry University}
  \city{Coventry}
  \country{U.K.}
}

\author{David Croft}
\email{David.Croft@coventry.ac.uk}
\affiliation{%
  \institution{Coventry University}
  \city{Coventry}
  \country{U.K.}
}


\begin{abstract}
  We ran a study on engagement and achievement for a first year undergraduate programming module which  used an online learning environment containing tasks which generate automated feedback.  Students could also access human feedback from traditional labs. We gathered quantitative data on engagement and achievement which allowed us to split the cohort into 6 groups.  We then ran interviews with students after the end of the module to produce qualitative data on perceptions of what feedback is, how useful it is, the uses made of it, and how it bears on engagement. A general finding was that human and automated feedback are different but complementary. However there are different feedback needs by group. Our findings imply: (1) that a blended human-automated feedback approach improves engagement; and (2) that this approach needs to be differentiated according to type of student. We give implications for the design of feedback for programming modules. 
\end{abstract}



\keywords{Feedback, Programming Education, Interactive Learning Environment}

\maketitle

\section{Introduction}
Student engagement is a major issue / goal in education. Student engagement is thought to improve achievement outcomes on one hand, and on the other, where engagement occurs thought to entify good teaching: therefore high engagement reflects both good teaching and good learning. Thus, engagement is a high priority. Engagement can be thought about as something students do (engage with learning), or teachers do (engage students in learning). Our aim is to treat feedback (plus other resources) as an engagement challenge - i.e. something we provide - which can evoke an engagement response as well as develop students' skills, satisfaction and outcomes.

The hypothesis driving this research is that feedback improves engagement and so feedback is a necessary part of good teaching and effective learning. 

The focus in this paper is on a situation where some feedback is automated-delivered (automated feedback from a learning environment) and other feedback is human - delivered by teaching staff face-to-face in co-located learning settings. Our interest is multi-fold: how these two forms of feedback differ, and how they are complementary. How are they experienced by students and what is their value? How do students get meaning out of feedback which improves their outcomes?

\section{Background}
\subsection{Engagement}
Engagement has been defined by \cite{trowler} as ``concerned with the interaction between the time, effort and other relevant resources invested by both students and their institutions intended to optimise the student experience and enhance the learning outcomes and development of students and the performance, and reputation of the institution''. This definition implies that students are partners in learning who make investment of time and resources to improve outcomes.

In our view engagement has three dimensions: first, behavioural, concerning e.g. attendance and time on task; second, cognitive, to do with students' understandings of their learning as well as metacognitive skills including reflection and self-regulation; and third, affective, to do with interest, and e.g. belonging. Importantly, this suggests that engagement is something both students and universities are involved in, with both parties contributing.

The importance on interactive engagement such as viewing, responding, changing or constructing is highlighted in \cite{Hosseini}. Engagement is defined as the extent of a student's active involvement in a learning activity. It can involve behavioural engagement, affective, cognitive and academic engagement \citep{Christenson}. Deep engagement improves learning. Authors of \cite{Hosseini} show that time spent on a learning task is positively correlated with learning and better performance. Engagement is enhanced with engaging programming examples. Learning from these engaging examples had a positive effect on student's engagement and learning.

\subsection{Feedback}
There are two types of feedback: formative and summative. Summative feedback is often comments on assessment scripts returned to students. Formative feedback can be around formative assessment or be ongoing: comments etc., written or spoken. \cite{Shute2008} defines formative feedback as ``information communicated to the learner that is intended to modify his or her thinking or behavior to improve learning''. There are various types: ``verification of response accuracy, explanation of the correct answer, hints, worked examples) and can be administered at various times during the learning process'' (e.g., immediately following an answer, after some time has elapsed). 

These characterizations of feedback reflect a constructivist perspective on learning \citep{Laurillard2002}, which assumes that learners are not passive receivers of information but agents who actively make sense of their own learning and seek meaning. On this view students can have different understandings of what learning is, with different experience and skills. This approach assumes that depending on this, feedback messages can be complex and need to be deciphered by students, rather than being something that can be assumed to straightforwardly be received and understood. 

The authors of \cite{Sergio} describe how learning to program evolves. It includes exploring different strategies and techniques in order to identify a programming solution. Support is needed during the learning process. Typically this support is provided by lecturers and teaching assistants. The support tries to help students to overcome misconceptions. We are interested to explore how students seek help in order to perform different programming tasks with a learning environment.

The author of \cite{sadler} proposes a `feedback gap'. Students need to have a concept of the goal of their learning; be able to compare their current performance with that goal; and take action to close the gap. However, this assumes that students must be able to self-assess in order to evaluate the gap and the actions that need to be taken; in turn, they should be able to make sense of and use feedback to improve performance. The implication of this is that feedback should support students in developing self-assessment skills in terms of closing the gap between performance and goal.

There has been research into types of feedback and effectiveness \citep{Shute2008}, often aimed at producing guidelines. Other research concerns students' response to and feelings about feedback. Authors of \cite{Higgins} show how students value feedback which promotes `deep learning' and critical thinking rather than `surface' feedback. \cite{Weaver} argues that the value of feedback depends on individual students' conceptions, depending on how far they share understandings of academic practice and discourse with tutors.

\section{Codio}
\begin{figure}[h]
  \centering
  \includegraphics[width=\linewidth]{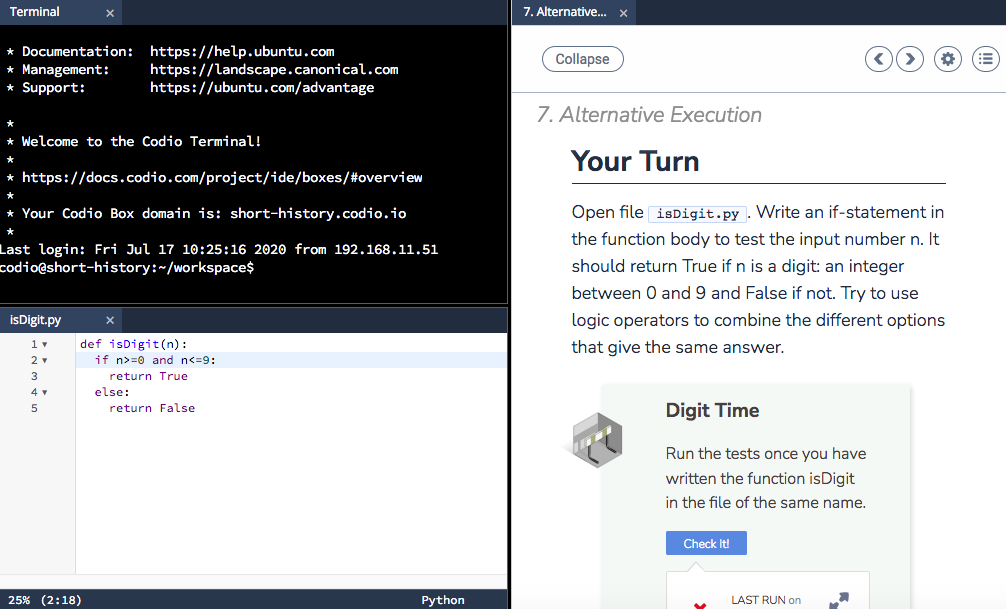}
  \caption{Codio learning environment.}
  \Description{The Codio learning environment.}
  \label{fig:codioTasks}
\end{figure}

As described by \cite{David} Codio is a learning platform where teachers prepare Codio units which include learning material and tasks that can provide formative feedback. Figure \ref{fig:codioTasks} shows a typical Codio unit. You can see the learning material on the right hand side and a program file with Linux terminal on the left. Students can enter their code and run it within the terminal. They can also use the check it button on the right to receive feedback.

\subsection{Automatic feedback and Codio}
\begin{figure}[h]
  \centering
  \includegraphics[width=\linewidth]{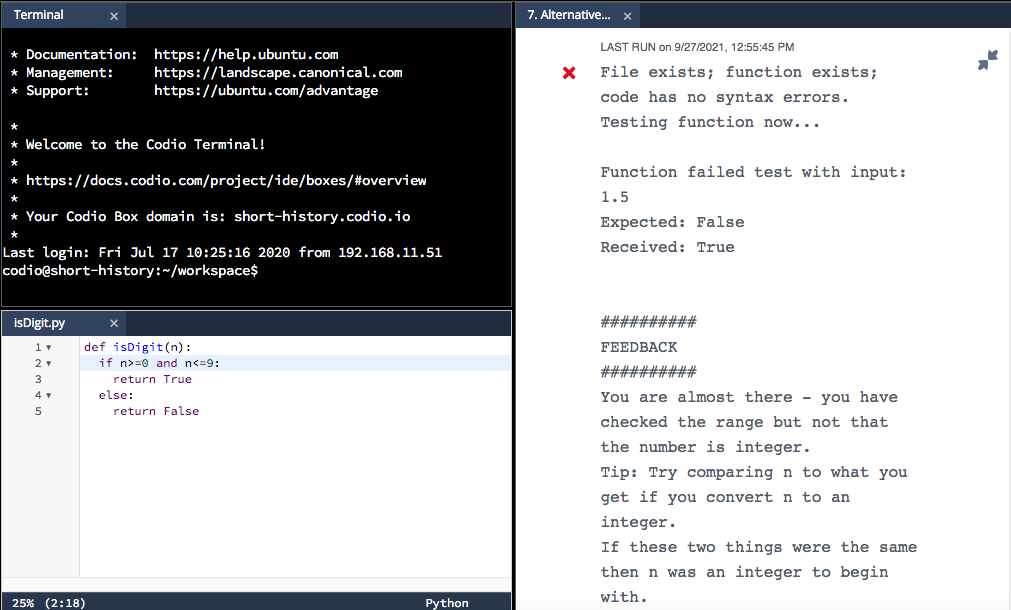}
  \caption{Formative feedback from within the Codio Unit.}
  \Description{Formative feedback from the Codio Unit.}
  \label{fig:CodioFeedback}
\end{figure}

Two different types of automatic formative feedback are available to students. First there is the feedback from the Linux terminal and second the feedback from the check it button inside the Codio unit. The feedback from within the Codio unit is developed by the instructor and is based on automated unit and system tests. Figure \ref{fig:CodioFeedback} shows typical feedback provided by the check it button. Students can check their code by clicking on the check it button. It will produce a green tick if the code is correct otherwise it will provide a red cross and some feedback if the code does not pass the tests.

\section{Method}
\subsection{Participants}
We interviewed 27 undergraduate computer science students. The students were in their first year and had just completed their Python programming module that used Codio.
\subsection{Study design}
This was a mixed-method study. There was automated data capture giving quantitative data on attendance (regarded as a broad - and flawed - index of engagement) and achievement (total marks for the module). This data was used to split students into 6 categories (see below).
Following the end of the module, we ran 20-minute interviews with students where we attempted a representative sample by split. The purpose of this was to gather qualitative data on students' experience of the module, particularly their perceptions on the nature and usefulness of human and automated feedback. A key focus was on whether and in what way these perceptions might differ by which of the 6 categories they were in and what are the further implications for the design and delivery of effective feedback.

\subsection{Category split}
The students were grouped into the following categories: Attendance was split into high (above 60\%) and low (below 60\%). Achievement was split into low (0-50\%), medium (50-70\%) and high (above 70\%).
Attendance was measured by students' swiping their ID as they enter the room (as used across the university). Achievement was the overall module score. 
Table \ref{tab:cat} shows the number of students per category.
\begin{table}
  \caption{Number of students per category}
  \label{tab:cat}
  \begin{tabular}{rccc}
    \toprule
    & &Achievement\\
    Attendance& low&medium&high\\
    \midrule
    low&  1 (LL)& 3 (LM)& 10 (LH)\\
    high& 2 (HL)& 4 (HM)&7 (HH)\\
  \bottomrule
\end{tabular}
\end{table}
\subsection{Interview design}
Interviews were run face-to-face with one student and two staff. These were run in quiet rooms, with audio recording, note-taking, and screen / person recording with screencapture, which captures interaction with Codio.

The structure of the interview was to ask what students thought about the module and allow interviews to follow student concerns. We wished to avoid leading questions on our research interests, but to progressively focus questions according to students' interests. Given this approach, a skeleton structure for the interview was: (1) broad questions; (2) detailed questions on issues based on student interests; and (3) an exploration of a Codio lesson chosen by the student where they talk through what they did and how they experienced it, with a focus on feedback and engagement.

This produced 27 datasets of both audio and video, which were analysed thematically, with an attempt to map themes to categories to expose differences; but also to see what themes are general.

\section{Findings and Analysis}
The quantitative data for the module allows a split (see Table \ref{tab:cat}). The table shows that the largest category was LH, i.e. low attendance and high achievement. This is followed by HH (high attendance and high achievement). The other 4 categories are smaller. LM means that low attendance students had medium achievement. This leaves the HM, high attendance and medium achievement group; HL, high attendance but low achievement; and LL: low attendance and low achievement. 


Students in the LH group all had previous programming experience.  3 out of the 10 students would not approach the lecturer as they felt they did not need help other than the automatic feedback; 4 reported that they would only approach the tutor if they did not understand the automatic feedback. Another 3 reported that they would ask a friend for help before approaching the tutor. 

The students in the LM group all had previous experience with programming. They did not have any problems to understand the automated feedback. One student reported that he would contact the tutor when they got stuck.

In contrast, the student in the LL group did not have any programming experience. The student said the attendance could have been better but felt sometimes demotivated. the student had problems to understand the automated feedback although the feedback from the tutor was perceived as more effective than the automatic feedback. However, as the student showed low attendance there was less experience of both automated and human feedback.

In the HH group 3 students reported that they did not have any previous programming experience. All students in this group made good use of the tutor feedback which they found very important as the tutor was able to explain the automatic feedback when it was difficult to understand. Interestingly nobody in this group reported that they would ask a friend for help. They asked the tutor directly when a problem was unclear.

All students in the HM group had previous experience with programming. 2 out of the 3 students reported that they would not ask the tutor for help as they were shy. One reported that they would ask a friend for help. 

Again, all students in the HL group had previous experience with programming. Both reported that they struggled at times with the programming tasks. They said that when there was a problem they got in touch with the tutor for help.

The LH group still engaged with the module without high attendance, through the remote use of the Codio lessons. These students were frequently of high ability due to previous experience and attainment, where the automated feedback on detailed issues was more interpretable than for lower ability (no prior experience and / or attainment) students where issues were more conceptual, to do with basic difficulties with fundamental programming concepts. These students were represented by the HL group, and the LL group. The HL group are of particular interest because there is an implication that feedback was less effective for students with low experience / attainment. 
\subsection{General themes}
Feedback was useful. Automated feedback was seen as different to human feedback. This shows expectations on what kind of feedback comes from each source, and where to look. 
Codio was seen as a useful (maybe essential) resource. It should be noted that this is despite the fact that the delivery method and interaction pattern for a lab session was on-Codio for at least 90 minutes, with lecturer feedback and solutions at the end. 

\subsection{Differences }
The HH group were able to talk clearly about the role and value of human and automated feedback. They made good use of the tutor in the programming sessions. 

The LH group also talked about peer support and peer feedback. Talking to friends about programming challenges was seen as useful, and could be a way of reducing the time asked of staff. For example some people said:
`If we can work it out on our own, there's no need to bother staff'.
This group experienced a rich set of feedback resources and were able to manage and sequence each type to achieve outcomes. 

Some students in the HM group reported that they were to shy to ask the tutor but also asked their friends for help, perhaps reflecting embarrassment about engaging with staff who taught this module. This group will not have experienced the valuable feedback from tutors but they found a way of dealing with problems by trying to resolve them via peer feedback.

In contrast the LM group showed less need than either HH or HL groups for feedback particularly peer-to-peer or staff. They gave nuanced accounts of how to use Codio feedback particularly error messages.
The HL group reported that they relied on tutor help in order to understand the automatic feedback. However this group did not seem to be able to make use of the feedback provided as effectively as the students in other groups.
The student in the LL group talked about issues regarding understanding Codio feedback. This student as well as the HL group were less able to either elicit or make use of feedback in the nuanced way the HH group did, implying that students are more or less expert not just in the domain but in how they learn and make use of feedback.



\section{Discussion and Conclusion}
All groups saw feedback as valuable, both human (staff) and automated. Therefore the issue is how to design feedback so that it engages effectively with all six types of students. 
The findings show that attendance and engagement are not the same thing. For the LH group, students can be engaged (via working on Codio at home) despite low attendance. Here the quality of automated feedback is paramount. Since Python interpreter feedback is a given, this needs no further attention, but unit testing becomes highly important. 
 
This is not true of the LL student, where there is a much greater need for basic conceptual guidance and arguably greater staff attention. They can be difficult to engage when they do attend, perhaps not wishing to be `bothered' by staff due to embarrassment about level and this being public. Some may be feedback resistant, i.e. not necessarily interested in their studies or engaging with staff. This is the most challenging group, where engagement faces basic issues with motivation and willingness. There may also be other issues about being comfortable in class, and feelings about self and their own abilities. 
 The HL group were better able to deal with the social nature of the experience but less able to act on feedback, which implies that its design and delivery were less useful for this group. The implications of this work are that a blended approach to feedback which recognises the usefulness and complementarity of different types is essential for learning programming. However there are clear differences between students with different feedback needs and different ability to make use of it. This suggests that 
Staff responsiveness, i.e. awaiting student requests, is not enough. While this works for HH, LH, HM and LM it is less appropriate for HL and LL where the approach needs to be proactive. 
This implies the nee for the development of tools to help staff be more aware of students' performance in class. But human feedback is often seen as something high-value and not to be sought until other attempts exhausted. However some students may not like the interaction and in this case it is important to develop kinds of automated feedback which could help this group. The difference seems to be about hints and the general shape of the program. 

There is also a need to recognise that the different groups are an outcome: what group students fall into is not necessarily predictable. This has implications for the value feedback can add; as well as where it is failing. This implies being able to track students in terms of how often they attend, how often they use feedback of what type, and how their use of, request for and experience of feedback may change. This also has a negative aspect where feedback is ineffective, and then stops, i.e. disengagement following attempts to engage. There are students who are disengaged from the outset and do not seek to study much if at all and these are the most difficult. 
A further important dimension is affective. It seems that low-achieving students are frequently confused and conflicted and may even dislike their experience and thus avoid it. This means that type and quality of feedback is not enough to consider: we also need to be aware of the affective context in which feedback is delivered \citep{umuai}. Thus feedback should engage not just with the cognitive but also the affective.
This moves into not just needing awareness tools of how students are doing cognitively but also affectively. There is evidence that affective state influences the cognitive, and this has implications for what use can be made of feedback, and what type of feedback should be given when a student is in a particular affective state. 
\bibliographystyle{ACM-Reference-Format}
\bibliography{ref-CEP}


\end{document}